\documentclass[aps,prl,twocolumn,superscriptaddress]{revtex4}
\usepackage{graphicx,color,ulem}

\begin{document}

\title{Low energy spin waves and magnetic interactions in SrFe$_2$As$_2$}
\author{Jun Zhao}
\affiliation{Department of Physics and Astronomy, The University of Tennessee, Knoxville,
TN 37996-1200}
\author{Dao-Xin Yao}
\affiliation{Department of Physics, Purdue University, West Lafayette, IN 47907}
\author{Shiliang Li}
\affiliation{Department of Physics and Astronomy, The University of Tennessee, Knoxville,
TN 37996-1200}
\author{Tao Hong}
\affiliation{Neutron Scattering Science Division, Oak Ridge National Laboratory, Oak
Ridge, TN 37831-6393}
\author{Y. Chen}
\affiliation{NIST Center for Neutron Research, National Institute of Standards and
Technology,Gaithersburg, MD 20899-6102}
\author{S. Chang}
\affiliation{NIST Center for Neutron Research, National Institute of Standards and
Technology,Gaithersburg, MD 20899-6102}
\author{W. Ratcliff II}
\affiliation{NIST Center for Neutron Research, National Institute of Standards and
Technology,Gaithersburg, MD 20899-6102}
\author{J. W. Lynn}
\affiliation{NIST Center for Neutron Research, National Institute of Standards and
Technology,Gaithersburg, MD 20899-6102}
\author{H. A. Mook}
\affiliation{Neutron Scattering Science Division, Oak Ridge National Laboratory, Oak
Ridge, TN 37831-6393}
\author{G. F. Chen}
\affiliation{Institute of Physics, Chinese Academy of Sciences, P. O. Box 603, Beijing
100080, China}
\author{J. L. Luo}
\affiliation{Institute of Physics, Chinese Academy of Sciences, P. O. Box 603, Beijing
100080, China}
\author{N. L. Wang}
\affiliation{Institute of Physics, Chinese Academy of Sciences, P. O. Box 603, Beijing
100080, China}
\author{E. W. Carlson}
\affiliation{Department of Physics, Purdue University, West Lafayette, IN 47907}
\author{Jiangping Hu}
\affiliation{Department of Physics, Purdue University, West Lafayette, IN 47907}
\author{Pengcheng Dai}
\email{daip@ornl.gov}
\affiliation{Department of Physics and Astronomy, The University of Tennessee, Knoxville,
TN 37996-1200}
\affiliation{Neutron Scattering Science Division, Oak Ridge National Laboratory, Oak
Ridge, TN 37831-6393}

\begin{abstract}
We report inelastic neutron scattering studies of magnetic excitations in
antiferromagnetically ordered SrFe$_{2}$As$_{2}$ ($T_{N}=220$ K), the parent
compound of the FeAs-based superconductors. At low temperatures ($T=7$ K),
the spectrum of magnetic excitations $S(Q,\hbar \omega )$ consists of a
Bragg peak at the elastic position ($\hbar \omega =0$ meV), a spin gap ($
\Delta \leq 6.5$ meV), and sharp spin wave excitations at higher energies.
Based on the observed dispersion relation, we estimate the effective
magnetic exchange coupling using a Heisenberg model. On warming across $T_{N}
$, the low temperature spin gap rapidly closes, with weak critical
scattering and spin-spin correlations in the paramagnetic state. The antiferromagnetic order in SrFe$_{2}$As$
_{2}$ is therefore consistent with a first order phase transition, similar to the structural lattice distortion.
\end{abstract}

\maketitle




The parent compounds of the high-transition temperature (high-$T_{c}$)
copper oxides are simple antiferromagnetic (AF) Mott insulators \cite{palee}
characterized by a very strong nearest neighbor AF exchange coupling $J$ ($
>100$ meV) in the CuO$_{2}$ planes \cite{coldea}. When holes or electrons
are doped into the CuO$_{2}$ planes, the character of the ground state is
fundamentally altered from an AF insulator to a superconductor with
persistent short-range AF spin correlations (excitations) \cite{tranquada}.
In the case of FeAs-based superconductors such as $R$FeAsO$_{1-x}$F$_{x}$
(where $R=$ La, Nd, Sm, Pr,...) \cite{kamihara,chen,gfchen,zaren} and $
A_{1-x}B_{x}$Fe$_{2}$As$_{2}$ ($A=$Ba, Sr, Ca, $B=$ K, Cs, Na) \cite
{rotter,gfchen2,sasmal,gwu}, although the undoped parent compounds are also
long-range ordered antiferromagnets with a collinear spin structure as shown
in Fig. 1a \cite{cruz,zhao,mcguire,chen1,huang,zhao1,goldman}, much is
unknown about the magnetic exchange coupling responsible for such a spin
structure. For example, early theoretical studies suggested that LaFeAsO has
a spin-density-wave (SDW) instability \cite{dong,mazin}. As a consequence,
the AF spin structure in these materials arises from quasiparticle
excitations across electron-hole pockets in a nested Fermi surface \cite
{mazin2}, much like SDW antiferromagnetism in metallic chromium (Cr) \cite
{fawcett}. Alternatively, a Heisenberg magnetic exchange model\cite
{yildirim,ma1,fang} is suggested to explain the AF structure. Here, the
collinear spin phase is stable when the nearest neighbor exchange $J_{1}$
and the next nearest neighbor exchange $J_{2}$ satisfy $J_{1}<2J_{2}$ (Fig.
1a). First-principles calculations estimate $J_{1}\sim J_{2}$\cite{ma1}. In
contrast, some band structure calculations\cite{yin} suggest that the $J_{1}$
along the $a$-axis and $b$-axis of the low temperature orthorhombic
structure ($c>a>b$) can have different signs with $J_{1a}$ and $J_{1b}$
being AF and ferromagnetic, respectively, and that $J_{1a}>2J_{2}$.
Therefore, there is no theoretical consensus on the relative strengths of $
J_{1a}$, $J_{1b}$, and $J_{2}$ or the microscopic origin of the observed AF
spin structure. If magnetism is important for superconductivity of these
materials, it is essential to establish the \textquotedblleft effective
Hamiltonian\textquotedblright\ that can determine the magnetic exchange
coupling.

In this paper, we report inelastic neutron scattering studies of spin wave
excitations in single crystals of SrFe$_{2}$As$_{2}$ ($T_{N}=220$ K) \cite
{zhao1}. At low temperature, we find that spin waves have an
anisotropy gap of $\Delta =6.5$ meV and disperse rapidly along both the $
[H,0,0]$ and $[0,0,L]$ directions. On warming to 160 K, the magnitude of the
spin gap decreases to 3.5 meV while the intensities of the spin wave
excitations follow the expected Bose statistics. However, there are only weak 
critical scattering and magnetic correlations in the paramagnetic
state at 240 K, in sharp contrast to the SDW excitations in Cr \cite{fawcett}
and spin waves in cuprates \cite{coldea,tranquada}. We estimate the
effective magnetic exchange coupling using a Heisenberg model and find that $
J_{1a}+2J_{2}=100\pm 20$ meV, $J_{z}=5\pm 1$ meV, with a magnetic single ion
anisotropy $J_{s}=0.015\pm 0.005$ meV. The weak critical scattering and
paramagnetic spin-spin correlations, together with the simultaneous 
first order structural transition \cite{zhao1}, suggest a first 
order AF phase transition.

Our experiments were carried out on HB-1 triple-axis spectrometer at the
High Flux Isotope Reactor, Oak Ridge National Laboratory, BT-7 and SPINS
triple-axis spectrometers at the NIST Center for Neutron Research. For HB-1
and BT-7 measurements, we fixed the final neutron energy at $E_{f}=14.7$ meV
and used PG(0,0,2) (pyrolytic graphite) as monochromator and analyzer. A PG
filter was placed in the exit beam path to eliminate $\lambda /2$. For the
SPINS measurements, the final neutron energy was fixed at $E_{f}=4.9$ meV
and a cold Be filter was placed in the scattered beam path. SrFe$_{2}$As$_{2}
$ single crystals were grown from flux \cite{gfchen2} and coaligned
within 2 degrees to have a total mass of $\sim $0.7 g. From earlier
diffraction work \cite{zhao1}, we know that the AF order occurs in close
proximity to the lattice distortion, changing the crystal structural
symmetry from tetragonal above $T_{N}$ to orthorhombic below it (Fig. 1b).
However, it is unclear whether the structural and magnetic phase transitions
are second \cite{tegel} or first order \cite{jesche}. For the observed spin
structure (Fig. 1a), magnetic Bragg peaks are allowed at $[H,0,L]$ ($H,L$
are odd integers) reciprocal lattice units (r.l.u), where the momentum
transfer is $Q$(in \AA $^{-1})=(H2\pi /a,K2\pi /b,L2\pi /c)$ and $a=5.5695(9)
$, $b=5.512(1)$, $c=12.298(1)$ \AA\ are lattice parameters in the
orthorhombic state at 150 K. To probe spin wave excitations, we aligned our
single crystal array in the $[H,0,L]$ zone, where we can probe excitations
along the $[H,0,0]$ and $[0,0,L]$ directions.

\begin{figure}[tbp]
\includegraphics[scale=.35]{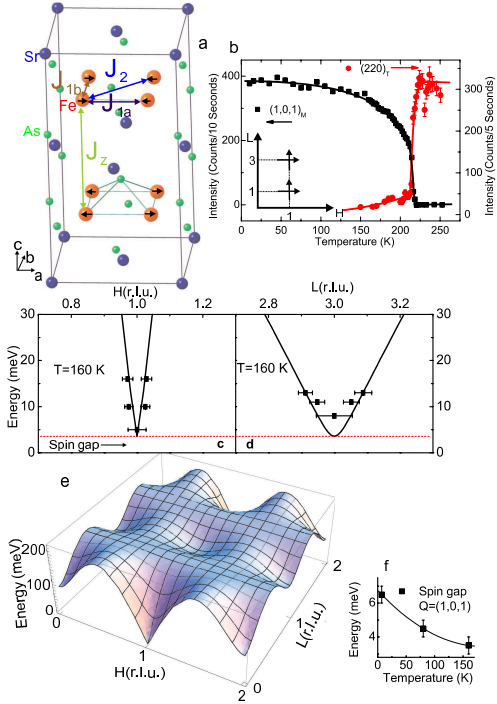}
\caption{(Color online) (a) The Fe spin ordering in the SrFe$_{2}$As$_{2}$
chemical unit cell and magnetic exchange couplings along different
high-symmetry directions. (b) The AF N$\mathrm{\acute{e}}$el temperature and
the temperature dependence of the structural $(2,2,0)$ Bragg peak of the SrFe
$_{2}$Sr$_{2}$ crystals used in the experiment \protect\cite{zhao1}. The
inset shows positions in reciprocal space probed in the experiment. (c)
Observed spin wave dispersion along the $[H,0,0]$ direction at 160 K. (d)
Similar dispersion along the $[0,0,L]$ direction. (e) Calculated
three-dimensional spin wave dispersions using $J_{1a}=20$, $J_{1b}=10$, $J_2=40$,
$J_z=5$, and $J_s=0.015$ meV. (f) Temperature dependence of the
anisotropy spin gap $\Delta (T)$.}
\end{figure}

\begin{figure}[tbp]
\includegraphics[scale=0.8]{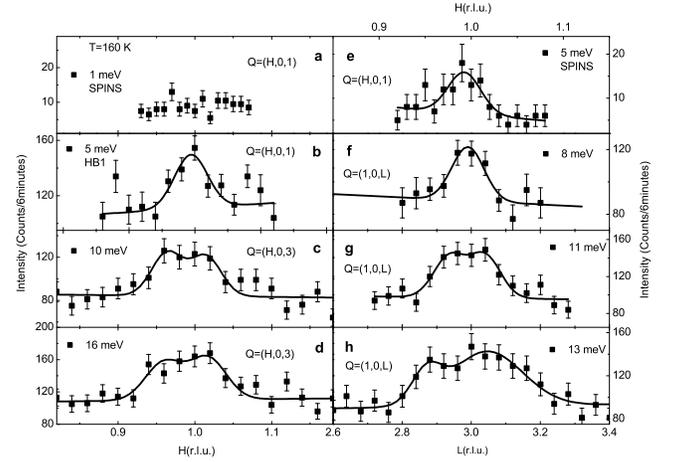}
\caption{Wave vector dependence of the spin wave excitations at 160 K
obtained on cold [(a) and (e)] and thermal [(b)-(d) and (f)-(h)] triple-axis
spectrometers at different energies. (a) $Q$-scan along the $[H,0,1]$
direction at $\hbar \protect\omega =1$ meV using SPINS. The spectrum is
featureless indicating the presence of a spin gap exceeding 1 meV. Identical scan at $
\hbar \protect\omega =5$ meV in (e) shows clear evidence of spin wave
excitations centered at $(1,0,1)$. (b-d) $Q$-scans along the $[H,0,1]$ or $
[H,0,3]$ directions at different energies. The spectra clearly broaden with
increasing energy. (f-h) Similar scans along the $[1,0,L]$ direction, which
probe the exchange coupling $J_{z}$.}
\end{figure}

Figures 2a and 2e show constant-energy scans for $\hbar \omega =1$ and 5 meV
around $[H,0,1]$ at 160 K obtained on SPINS. While the scattering at $\hbar
\omega =1$ meV is featureless (Fig. 2a), there is a clear peak centered at $
H=1$ in the 5 meV data (Fig. 2e). This immediately suggests that spin waves
in SrFe$_{2}$As$_{2}$ have an anisotropy gap at this temperature that is
less than 5 meV. Moving on to higher energies, Figures 2b-d and 2f-h
summarize $Q$-scans along $[H,0,0]$ and $[0,0,L]$ directions, respectively,
at different energies. The $Q$-widths of the scattering clearly become
broader with increasing energy. Figures 1c and 1d show the observed
dispersion curves for the limited energy range with observable spin wave
excitations. Assuming an effective Heisenberg Hamiltonian \cite{fang,yao} $
H=J_{1a}\sum_{i,j}\mathbf{S}_{i}\cdot \mathbf{S}_{j}+J_{1b}\sum_{i,j}\mathbf{
S}_{i}\cdot \mathbf{S}_{j}+J_{2}\sum_{i,j}\mathbf{S}_{i}\cdot \mathbf{S}
_{j}+J_{z}\mathbf{S}_{i}\cdot \mathbf{S}_{j}-J_{s}(S_{i}^{z})^{2}$, where $
J_{1a}$, $J_{1b}$, $J_{2}$, and $J_{z}$ are exchange interactions shown in
Fig. 1a; $J_{s}$ is the single ion anisotropy; and $S$ is the magnitude of
iron spin, the spin wave dispersions along the $[H,0,0]$ and $[0,0,L]$
directions near the $(1,0,1)$ Bragg peak are $
E(k_{x})=2S[(J_{1a}+2J_{2}+J_{s}+J_{z})^{2}-(J_{z}-(J_{1a}+2J_{2})\cos {k_{x}
})^{2}]^{1/2}$ and $E(k_{z})=2S[(2J_{1a}+4J_{2}+J_{s}+J_{z}-J_{z}\cos {k_{z}}
)(J_{s}+J_{z}+J_{z}\cos {kz})]^{1/2}$, respectively. In addition, the size
of the spin gap due to the single ion anisotropy is $\Delta
(1,0,1)=2S[J_{s}(2J_{1a}+4J_{2}+J_{s}+2J_{z})]^{1/2}$. The solid lines in
Figs. 1c and 1d are the best fits with these equations, where $
J_{1a}+2J_{2}=100\pm 20$ meV, $J_{z}=5\pm 1$ meV, and $J_{s}=0.015\pm 0.005$
meV. The three-dimensional plot in Fig. 1e shows the expected spin wave
dispersion at higher energies.

To determine the temperature dependence of the spin gap, we carried out
energy scans at the signal and background positions for spin waves at
different temperatures. At 7 K, an energy scan at the magnetic zone center position $
Q=(1,0,1)$ shows an abrupt increase above 6.5 meV, while the background
scattering at $Q=(1.2,0,1)$ is featureless (Fig. 3a). Energy scans at
equivalent positions $Q=(1,0,3)$ and $(0.8,0,3)$ in Fig. 3b show similar
results and therefore reveal a low temperature spin gap of $\Delta =6.5$
meV. On warming to 80 K, identical scans at $Q=(1,0,1)$ and $Q=(1.2,0,1)$
show that the spin gap is now at $\Delta =4.5$ meV (Fig. 3c). Finally, $
\Delta $ becomes 3.5 meV at 160 K, consistent with constant-energy scans in
Figs. 2a and 2e. These results indicate that the spin anisotropy of the
system reduces with increasing temperature.

\begin{figure}[tbp]
\includegraphics[scale=0.75]{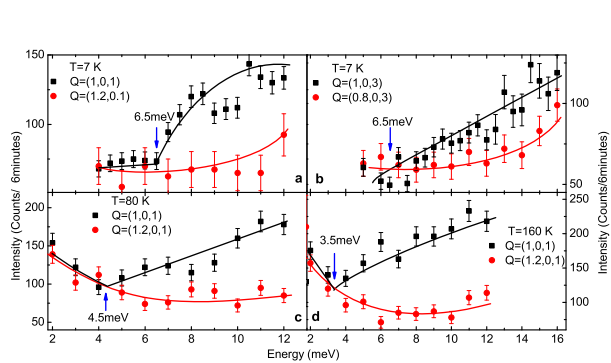}
\caption{(Color online) Temperature dependence of the spin-gap obtained from
energy scans around the (1,0,1) and (1,0,3) Bragg peaks. (a) Low temperature
($T=7$ K) constant-$Q$ scans at the signal ($Q=(1,0,1)$) and background ($
Q=(1.2,0,1)$) positions show a clear spin-gap of $\Delta =6.5$ meV (b)
Similar scans at $Q=(1,0,3)$ and $Q=(0.8,0,3)$ which again show $\Delta =6.5$
meV. (c),(d) Temperature dependence of the spin-gap, where $\Delta =4.5$ meV
at 80 K and $\Delta =3.5$ meV at 160 K.}
\end{figure}

\begin{figure}[tbp]
\includegraphics[scale=.6]{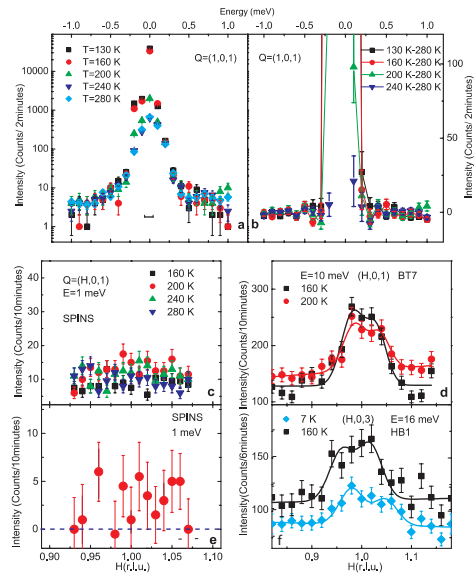}
\caption{(Color online) Temperature dependence of the quasi-elastic magnetic
scattering and spin wave excitations below and above $T_{N}$. Data in
(a)-(c) are obtained on SPINS. (a) Constant-$Q$ scans at the  $Q=(1,0,1)$
Bragg peak position at different temperatures. Except for the dramatic
increase in the elastic component below $T_{N}$, the quasi-elastic scattering above $\hbar \protect
\omega =0.5$ meV is essentially temperature independent, revealing no 
evidence for the Lorentzian-like paramagnetic scattering above $T_{N}$
observed in Cr \protect\cite{fawcett}. (b) Temperature difference spectra
using $T=280$ K scattering as background. The data again show little evidence of
critical scattering at $\hbar\omega=1$ meV. (c) $Q$
-scans at $\hbar \protect\omega =1$ meV at different temperatures. We speculate that
the slight increase in overall scattering at 240 K from 280 K shown in (e)
is due to weakly correlated paramagnetic spins. (d) $\hbar \protect\omega =10
$ meV spin-wave excitations at 160 K and 200 K obtained on BT-7. (f) $\hbar 
\protect\omega =16$ meV spin-wave excitations at 7 K and 160 obtained on
HB-1. The intensity increase is due to the Bose population factor.}
\end{figure}

Figure 4 summarizes the temperature dependence of the spin waves and
quasielastic scattering in the AF ordered and paramagnetic states obtained
on SPINS and BT-7. Figure 4a shows energy scans at $Q=(1,0,1)$ for $
T=130,160,200,240,$ and 280 K plotted on a log scale. When the temperature is
increased across $T_{N}$ ($=220$ K), there is a rapid decrease in the
ordered moment but little evidence for quasielastic and critical scattering,
which are signatures of a second order phase transition. To illustrate this
point, we plot in Fig. 4b the temperature difference scattering using 280 K
data as background. Besides the magnetic Bragg peak below $T_{N}$ at $\hbar
\omega =0$ meV, there is little quasielastic critical scattering
typical of a second order phase transition. Figure 4c shows constant-energy
scans ($\hbar \omega =1$ meV) measured on SPINS and the scattering is
essentially featureless at all temperatures investigated. 
Assuming no spin correlations in the paramagnetic
state at 280 K, the differences in the scattering between 240 K and 280 K at $\hbar\omega=1$ meV
should reveal the magnetic intensity gain close to $T_{N}$. Consistent with the 
temperature dependence of the energy scans in Figs. 4a and 4b, there are
signs of possible uncorrelated paramagnetic scattering (since the subtracted
data in Fig. 4e are overall positive) at 240 K but weak critical
scattering. For temperatures below $T_{N}$, we find that spin wave
excitations at temperatures below 160 K simply follow the Bose statistics
(Fig. 4f). 

The discovery of the collinear AF order with small moment in LaFeAsO \cite
{cruz} has caused much debate about its microscopic origin. Since LaFeAsO is
a semimetal, the observed AF order may arise from a SDW instability due to
Fermi surface nesting \cite{dong,mazin,mazin2}, where electron itinerancy is
important much like incommensurate SDW order in pure metal Cr \cite{fawcett}
. Alternatively, there are reasons to believe that LaFeAsO is in proximity
to a Mott insulator \cite{si}, and the AF order is a signature of local
physics and electron correlations \cite{fang,xu}. Another heavily debated
issue is the first \cite{jesche} or second \cite{tegel} order nature of the
simultaneous structural/magnetic phase transition in SrFe$_2$As$_2$.

If the observed AF order in SrFe$_{2}$As$_{2}$ originates from Fermi surface
nesting similar to the SDW order in Cr, the velocity of the spin waves $c$
should be $c=\sqrt{v_{e}v_{h}/3}$, where $v_{e}$ and $v_{h}$ are the
electron and hole Fermi velocities, respectively \cite{fawcett}. The
dispersion relation is then $\hbar \omega =cq$ where $q$ is the magnitude of
the momentum transfer away from the Bragg position. For Cr, the spin-wave
velocity is measured to be $c=851\pm 98$ meV-\AA \cite{fawcett}. In
addition, there are strong spin-spin correlations in the paramagnetic state
where the dynamic structure factor $S(q,\hbar \omega )$ can be described by
the product of a Gaussian centered at the SDW ordering wave vector and a
Lorentzian in energy, or $S(q,\hbar \omega )=S_{0}(T)e^{-\xi /2\sigma
^{2}}(\hbar \omega /k_{B}T)/[((\hbar \omega )^{2}+\Gamma )(1-e^{-\hbar
\omega /k_{B}T})]$, where $\sigma $ and $\Gamma $ are the Gaussian and
Lorentzian widths, respectively \cite{fawcett}. At temperatures as high as
500 K ($T=1.6T_{N}$), one can observe a clear resolution-broadened
Lorentzian centered at $\hbar \omega =0$ meV with $\Gamma =15.6$ meV \cite
{fawcett}. For comparison, there is no evidence of a Lorentzian-like
quasielastic scattering in SrFe$_{2}$As$_{2}$ even at $T=1.09T_{N}$. The
lack of critical scattering both below and above $T_{N}$, together with the
fact that there is also an abrupt structural distortion occurring at the same
temperature and observed thermal hysteresis \cite{zhao1,goldman,tegel,jesche}, is consistent with the AF phase transition being
first order in nature.

To compare the observed exchange couplings in Fig. 1 and those expected from
SDW excitations in a nested Fermi surface, we note that Fermi velocities
estimated from the local density approximation calculations for BaFe$_{2}$As$%
_{2}$ \cite{ma2} are $v_{e}=2.2$ eV-\AA\ and $v_{h}=1.2$ eV-\AA . Assuming
BaFe$_{2}$As$_{2}$ and SrFe$_{2}$As$_{2}$ have similar Fermi velocities, the
expected spin wave velocity is then $c\sim 0.94$ eV-\AA . However, since
Angle Resolved Photoemission spectroscopy (ARPES) experiments on BaFe$_{2}$As%
$_{2}$ \cite{yang} show that the band-width is strongly renormalized, the
larger Fermi velocities in electron and hole pockets are $v_{e}\approx
v_{h}\sim 0.5$ eV-\AA . These values would give $c\sim 0.29$ eV-\AA . Using
smaller Fermi velocities would yield half of the larger values or $c\sim 0.15
$ eV-\AA . Within the local moment effective $J_{1a}-J_{1b}-J_{2}$ model,
the spin wave velocity is given by $c=(J_{1a}+2J_{2})a/2$ eV-\AA . From our
measured $J_{1a}+2J_{2}=100\pm 20$ meV, $c\sim 0.28$ eV-\AA\ which is also fairly
close to the ARPES results. Therefore, our present data do not allow an
unambiguous distinction between localized and itinerant description of the
AF order in SrFe$_{2}$As$_{2}$ in terms of the spin wave velocity.

In summary, we carried inelastic neutron scattering experiments to study low
energy spin wave excitations in SrFe$_2$As$_2$. The low-temperature spectrum
consists of a Bragg peak, a spin gap, and sharp spin wave excitations at
higher energies. Using a simple Heisenberg Hamiltonian, we find $
J_{1a}+2J_2=100\pm 20$ meV, $J_z=5\pm 1$ meV, and $J_s=0.015\pm0.005$ meV.
On warming across $T_N$, there is weak critical scattering and
spin-spin correlations in the AF wave vector region explored 
in the paramagnetic state, different from the
paramagnetic SDW excitations in Cr. These results are consistent with the AF phase
transition in SrFe$_2$As$_2$ being first order in nature.

We thank R. Fishman for discussions on Cr. This work is supported by the US NSF DMR-0756568, by NSF PHY-0603759, by the
US DOE, BES, through DOE DE-FG02-05ER46202 and Division of Scientific User
Facilities. The work at the Institute of Physics, Chinese Academy of
Sciences, is supported by the NSF of China, the Chinese Academy of Sciences
and the Ministry of Science and Technology of China.


\end{document}